# Frequency tuning capability of self-powered piezoelectric tuned mass dampers

**D. Gibus, O. Larhlid, A. Guerine, A. Badel**
Université Savoie Mont-Blanc, SYMME
F-74000 Annecy, France,
e-mail: **david.gibus@univ-smb.fr**

## Abstract

Tuned mass dampers suffer from frequency shift sensitivity. This paper investigates the use of tunable piezoelectric tuned mass dampers (PE-TMD) by studying their behavior with tunable resistive and capacitive loads to adapt their resonant frequency. An analytical model is developed to assess the influence of key dimensionless parameters—electromechanical coupling coefficient, mass ratio, and quality factor—on vibration attenuation, frequency tuning capability, and energy harvesting. The study reveals trade-offs between attenuation, tunability, and energy harvesting. It provides design guidelines for autonomous PE-TMDs and identifies key challenges for their practical implementation, including the interest of high coupling coefficients, whose the implication is limited is with growing mass ratio. Notably, it highlights the necessity of high coupling coefficient, the influence of which on frequency tuning is limited as the mass ratio grows.

## 1 Introduction

The use of tuned mass dampers (TMD) is well known for the attenuation of structural vibrations. They are tuned to have a natural frequency close to that of the host structure, enabling the attenuation of vibrations within a limited frequency range. One of the main issues hindering performance is their sensitivity to frequency shifts [1]. Recent literature has demonstrated an increased level of interest in electromechanical systems, known as semi-active TMDs, that are able to adapt to the behavior of TMDs [2]. As these systems require energy for vibration control, they face significant challenges regarding their integration and maintenance. External energy is necessary to emulate electronic components, which would be too complex to realize in practice (e.g., negative capacitances [3] or inductances with low series resistance [4]).

On the other hand, in the field of vibration energy harvesting it has been shown that the resonant frequency and equivalent damping of piezoelectric vibration harvesters can be controlled using self-powered electronic circuits [5]. These semi-passive interfaces involve only transistors with controlled conduction, where the semi-active interface involves the use of energy-consuming circuits with operational amplifiers [4]. However, the influence of the electromechanical coupling coefficient ($k_e^2$) and the TMD's structural parameters on the frequency tuning capability has not been thoroughly investigated. In this work, we propose to theoretically study the performance of piezoelectric tuned mass dampers (PE-TMD).

Semi-passive circuits can be analyzed, at the first harmonic, as equivalent to a passive circuit [6]. While resistive and reactive loads can be studied, inductive loads with low resistive losses are complex to emulate in an autonomous circuit [7], [8], [9]. Therefore, in order to provide an initial analysis of the possibilities offered by autonomous circuits, we focus here the study on variable parallel capacitive and resistive loads. This work provides an initial insight of the challenges associated with the electrical tuning of the resonance frequency of PE-TMDs, prior to conducting a more in-depth study on the use of adaptive semi-passive circuits (e.g., the adaptive SSDI [10]) in PE-TMDs.

We examine the influence of dimensionless parameters—namely, the mass ratio, the electromechanical coupling coefficient, and the quality factor—on the vibration attenuation of the primary structure, the frequency tuning capability, and the harvested energy. The model is presented in the following section. Frequency responses are then presented in third section in order to show the influence of the electrical loads. The fourth section is devoted to parametric studies. Experimental results will be presented at the conference.

# 2 Modelling

## 2.1 Device and hypotheses

The system under study consists of a primary structure (mass $M_1$, stiffness $K_1$, damping $C_1$) coupled with a PE-TMD (mass $M_2$, short circuit stiffness $K_2$, damping $C_2$, piezoelectric capacitance $C_{\mathrm{p}}$, and electromechanical coupling term $\alpha$). An external force $f_e$ is applied to the mass $M_1$. The PE-TMD is connected to an equivalent impedance load Z, that correspond to a resistive load and a capacitive load connected in parallel. The schematic diagram of the lumped model is illustrated in Figure 1.

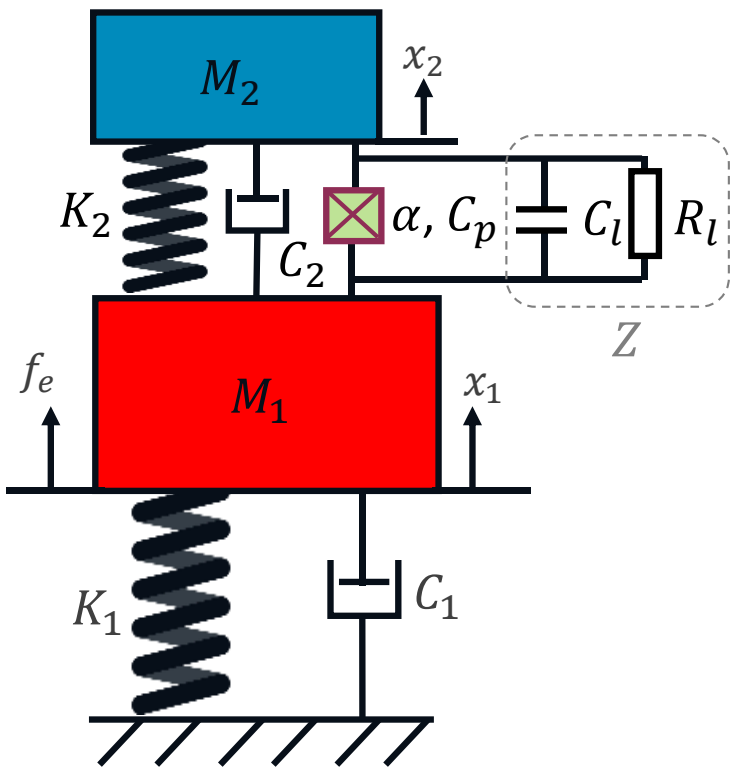

Figure 1 : Schematic diagram of the lumped model of primary structure with the PE-TMD.

The equations describing the behavior of the primary structure with PE-TMD can be found in [11]. The equation of motion is given in (1) and the equation of the current is given in (2). $x_1$ and $x_2$ are the displacement of masses $M_1$ and $M_2$ respectively. $v$ is the voltage across the electrodes of the piezoelectric element.

$$\begin{bmatrix} M_1 & 0 \\ 0 & M_2 \end{bmatrix}\begin{bmatrix} \ddot{x_1} \\ \ddot{x_2} \end{bmatrix} + \begin{bmatrix} C_1 + C_2 & -C_2 \\ -C_2 & C_2 \end{bmatrix}\begin{bmatrix} \dot{x_1} \\ \dot{x_2} \end{bmatrix} + \begin{bmatrix} K_1 + K_2 & -K_2 \\ -K_2 & K_2 \end{bmatrix}\begin{bmatrix} x_1 \\ x_2 \end{bmatrix} + \begin{bmatrix} -\alpha \\ \alpha \end{bmatrix} v = \begin{bmatrix} f_e \\ 0 \end{bmatrix} \quad (1)$$

$$i = -C_p\,\dot{v} + \alpha(\dot{x}_2 - \dot{x}_1) \quad (2)$$

For sinusoidal external excitation at an angular frequency ω expressed as in $\widetilde{f_e}(\omega) = Fe^{j\omega t}$, Eq. (2) can be reduced to Eq. (3), by inserting the corresponding frequency responses $\widetilde{x_1}(\omega) = \widetilde{X_1}e^{j\omega t}$ , $\widetilde{x_2}(\omega) = \widetilde{X_2}e^{j\omega t}$ and $\tilde{v}(\omega) = \tilde{V}e^{j\omega t}$. $\tilde{Z}$ is a complex impedance that is a function of $\omega$ as it corresponds to a capacitive load $C_l$ and a resistive load in parallel $R_l$ as given in (4).

$$\tilde{V} = \frac{j\omega\alpha\tilde{Z}}{1+j\omega C_p\tilde{Z}}\left(\widetilde{X_2} - \widetilde{X_1}\right) \quad (3)$$

$$\tilde{Z} = \frac{R_l}{R_l C_l j\,\omega + 1} \quad (4)$$

Replacing $\tilde{v}$ by its expression from (3) in (1) leads to equation (5).

$$\left\{-\omega^2\begin{bmatrix} M_1 & 0 \\ 0 & M_2 \end{bmatrix} + j\omega\begin{bmatrix} C_1 + C_2 & -C_2 \\ -C_2 & C_2 \end{bmatrix} + \begin{bmatrix} K_1 + K_2 & -K_2 \\ -K_2 & K_2 \end{bmatrix} + \frac{j\omega\alpha^2\tilde{Z}}{1+j\omega C_p\tilde{Z}}\begin{bmatrix} 1 & -1 \\ -1 & 1 \end{bmatrix}\right\}\begin{bmatrix} \widetilde{X_1} \\ \widetilde{X_2} \end{bmatrix} = \begin{bmatrix} F \\ 0 \end{bmatrix} \quad (5)$$

To generalize the analysis, the system is normalized in (6) using the dimensionless parameters from (7). $\omega_1$ is the natural pulsation of the primary structure alone, $\omega_2$ is the natural pulsation of the PE-TMD alone at short-circuit condition, $Q_{m1}$ and $Q_{m2}$ are the quality factors of the primary structure and PE-TMD respectively, $k_e$ is the expedient electromechanical coupling coefficient of the PE-TMD.

$$\left\{-\frac{\omega^2}{\omega_1^2}\begin{bmatrix}1 & 0\\ 0 & \mu\end{bmatrix}+\frac{j\omega}{\omega_1}\begin{bmatrix}\frac{1}{Q_{m_1}}+\mu\frac{\omega_2}{\omega_1}\frac{1}{Q_{m_2}} & -\mu\frac{\omega_2}{\omega_1}\frac{1}{Q_{m_2}}\\ -\mu\frac{\omega_2}{\omega_1}\frac{1}{Q_{m_2}} & \frac{1}{Q_{m_1}}\end{bmatrix}+\begin{bmatrix}1+\mu\left(\frac{\omega_2}{\omega_1}\right)^2 & -\mu\left(\frac{\omega_2}{\omega_1}\right)^2\\ -\mu\left(\frac{\omega_2}{\omega_1}\right)^2 & 1\end{bmatrix}+\left(\frac{\omega_2}{\omega_1}\right)^2\frac{1}{\mu}k_e^2\frac{j\frac{\omega}{\omega_2}\tilde{Z}}{1+j\frac{\omega}{\omega_2}\tilde{Z}}\begin{bmatrix}1 & -1\\ -1 & 1\end{bmatrix}\right\}\begin{bmatrix}\widetilde{X_1}\\ \widetilde{X_2}\end{bmatrix}=\frac{1}{K_1}\begin{bmatrix}F\\ 0\end{bmatrix} \quad (6)$$

$$\omega_1=\sqrt{\frac{K_1}{M_1}},\omega_2=\sqrt{\frac{K_2}{M_2}},\mu=\frac{M_2}{M_1},Q_{m1}=\frac{M_1\omega_1}{C_1},Q_{m2}=\frac{M_2\omega_2}{C_2},k_e^2=\frac{\alpha^2}{K_2C_p} \quad (7)$$

This normalization allows for a generalized analysis of the system's behavior, independent of specific physical dimensions. In this work, the velocity of the primary structure $\dot{x_1}$ is studied using the transfer function as $T_1$ expressed in (8).

$$T_1(\omega)=\frac{j\omega\widetilde{X_1}}{F} \quad (8)$$

This study aims to minimize the peak amplitude of the primary structure's velocity as realized in [12]. The resistive and capacitive loads are optimally selected to achieve effective vibration attenuation.

# 3 Response of the PE-TMD

The parametric study is conducted using the values summarized in Table 1. The resonant pulsation ($\omega_1$), primary structure mass ($M_1$), and piezoelectric capacitance ($C_p$) are set to unity for simplicity, while the quality factors ($Q_{m1}, Q_{m2}$) are fixed at 90 as it is a common value for structures and piezoelectric devices. The mass ratio (μ) and coupling coefficient ($k_e^2$) are varied to investigate their influence on the system's performance.

Table 1: Parameters of the study

| Name | Symbol | Value |
|---|---|---|
| Resonant pulsation | $\omega_1$ | 1 rad/s |
| Primary structure mass | $M_1$ | 1 kg |
| Capacitance | $C_p$ | 1 µF |
| Primary structure quality factor | $Q_{m_1}$ | 90 |
| Piezoelectric device quality factor | $Q_{m_2}$ | 90 |
| Mass ratio | µ | 0.1 % to 10 % |
| Coupling coefficient | $k_e^2$ | 0.05 to 0.4 |

The transfer functions are computed for the primary structure's velocity and analyzed under three configurations: the primary structure alone (no TMD), the mechanical TMD ($k_e^2$ = 0, no piezoelectric effect), and the PE-TMD with optimized resistive ($R_l$) and capacitive ($C_l$) loads. The optimal values of $C_l$ and $R_l$ are determined using the *fminsearch* function of Matlab (which exploits the Nelder-Mead simplex algorithm) while ensuring that $R_l$ and $C_l$ remain positive, to not consider negative capacitance.

Figure 2.a. shows the response of the primary structure alone, the response using the TMD without the piezoelectric effect (i.e. $k_e^2 = 0$) for $\mu = 0.5$ % when optimizing $\omega_2$, and the response using optimized values of $C_l$ and $R_l$ for $k_e^2 = 0.2$. While the values are closes to the values provided in (9) and (10) by [13], they are determined numerically due to the presence of loss in the primary structure, and as the velocity is here in concern.

$$\left.\frac{\omega_2^{opt}}{\omega_1}\right|_{Q_{m1=\infty}}=\frac{1}{1+\mu}\sqrt{\frac{8\left[(4+3\mu)^{\frac{3}{2}}-\mu\right]\left(1+\sqrt{4+3\mu}\right)}{3(64+80\mu+27\mu^2)}} \quad (9)$$

$$\left.Q_{m2}^{opt}\right|_{Q_{m1}=\infty}=\frac{1}{2}\sqrt{\frac{6(1+\mu)^3\left(\sqrt{4+3\mu}-2\right)(64+80\mu+27\mu^2)}{\left(1+\sqrt{4+3\mu}\right)\left(2+3\mu-\sqrt{4+3\mu}\right)\left(64+88\mu+27\mu^2-8(4+3\mu)^{3/2}\right)}} \quad (10)$$

From Figure 2a., we observe that using convenient values $C_l$ and $R_l$ reduces the maximal amplitude of $T_1$ by inducing dissipation in the resistive load. It should be noted that $C_l$ is null; only a pure resistive load is sufficient to tune the PE-TMD and attenuate the vibration

Figure 2.b. represents the responses when the resonant frequency of the primary structure $\omega_1$ is modified by +5 % without changing $\omega_2$. $T_1$ is represented for the former impedance ($C_l = 0\ F$, and $R_l = 289$ kΩ) and for a recalculated optimal impedance ($R_l^{opt} = 1.02\ M\Omega$, $C_l^{opt} = 147$ nF). Figure 2.b. shows that a small variation of $\omega_1$ induces a variation of the maximum amplitude of $T_1$ if the impedance is not adjusted. Adjusting the impedance allows minimizing the amplitude. The points on Figure 2 represent the maximal amplitude of adjusted PE-TMD.

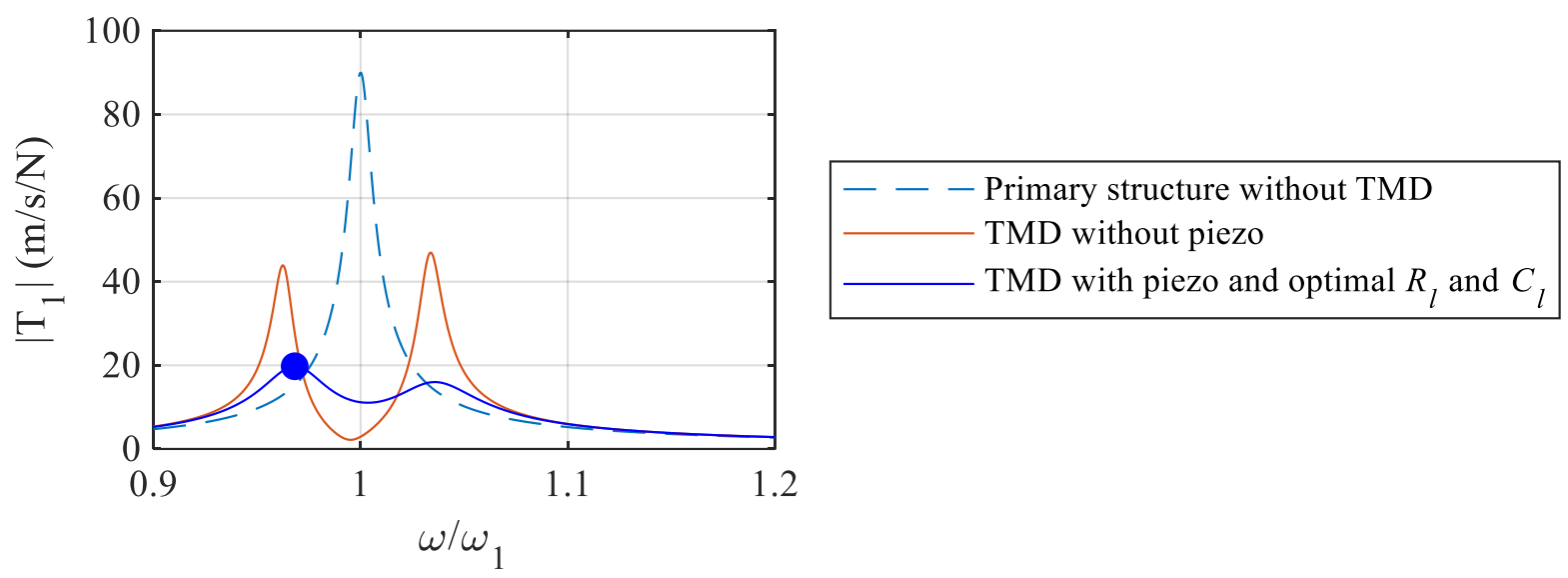

a.

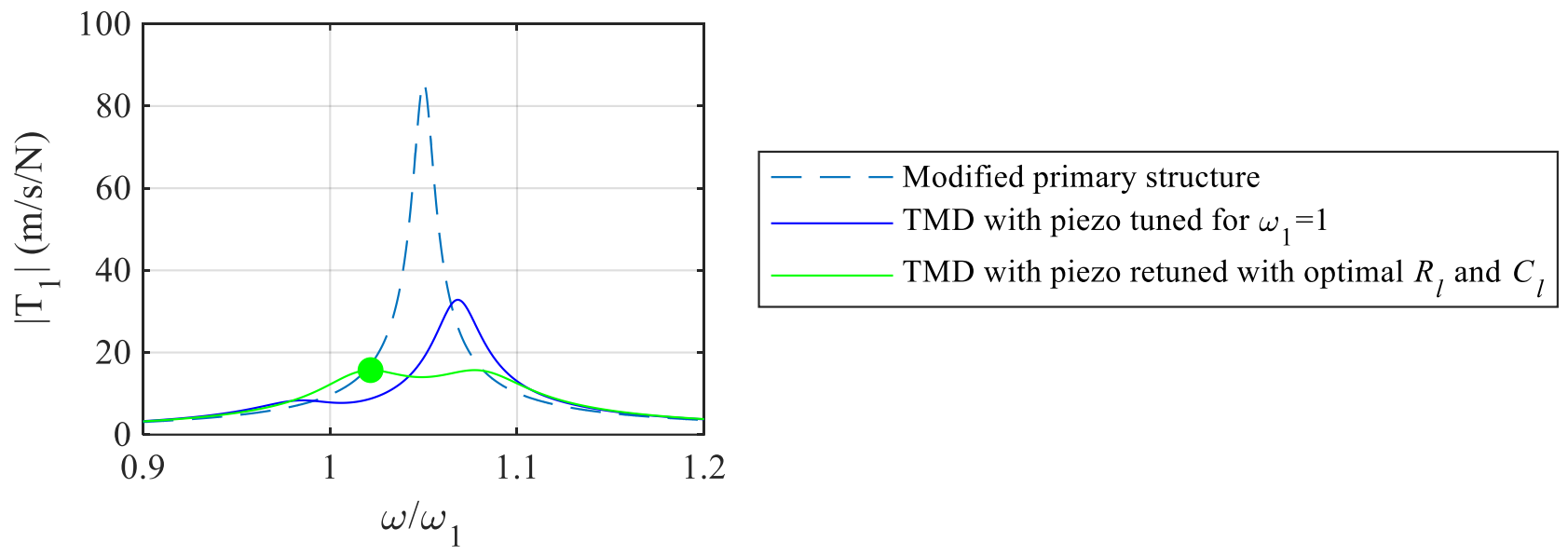

b.

Figure 2 : Transfer function amplitude $|T_1|$ a. for $\omega_1 = 1, C_l = 0$ and $R_l = 289\ k\Omega$, b. for $\omega_1 = 1.05$. The recalculated optimal loads are $R_l = 1.02\ M\Omega$, $C_l = 147\ nF$. μ = 0.5 % and $k_e^2 = 0.2$ for the responses with PE-TMD.

In order to study the influence of resistive and capacitive loads, Figure 3 shows the response of the PE-TMD with 10 values of $R_l$ and $C_l$ logarithmically spaced. $R_l$ varies between 0.1 MΩ and 10 MΩ with $C_l = 0\ F$ in Figure 3.a. and $C_l$ varies between 14.7 nF and 1.47 μF with $R_l = 0\ \Omega$ in Figure 3.b. It is observed that both modify the resonant frequency, by affecting the equivalent stiffness induced by the electromechanical coupling of the PE-TMD, as it can be observed in [6] for vibration energy harvesters. A pure resistive load $R_l$ affect the dissipation and the equivalent stiffness, while a pure capacitance only affects the equivalent stiffness. The combination of parallel resistive and capacitive loads affects both electrically induced stiffness and damping. Tuning $R_l$ and $C_l$ implies the possibility to tune the PE-TMD to optimal resonant frequency and equivalent damping. As passive loads are considered, the range of tunability is limited by the coupling of the $k_e^2$ as well as the mass ratio $\mu$ as discussed in section 4 a.
b.

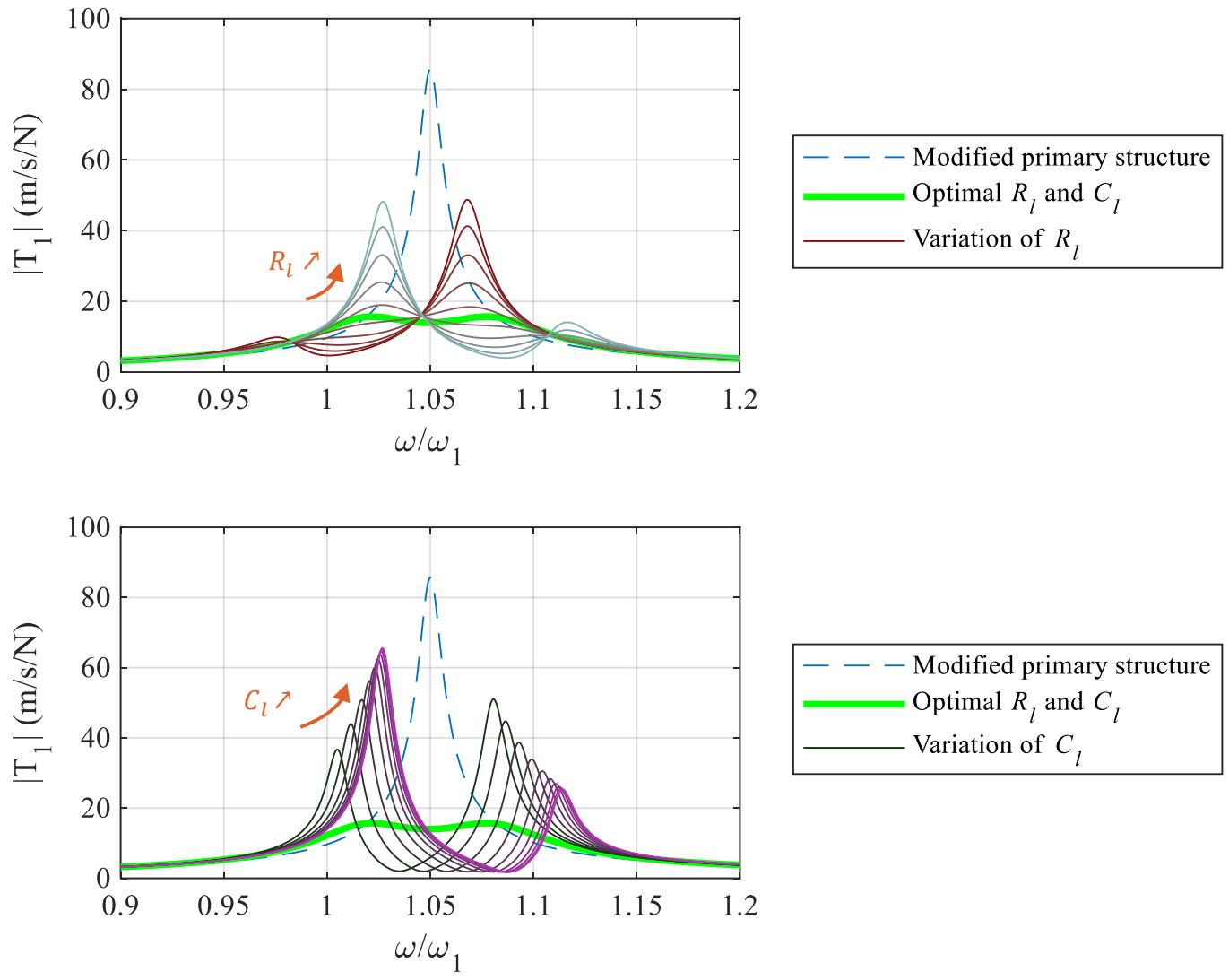

Figure 3 : Influence of the electrical loads on the $|T_1|$ for μ = 0.5 % and $k_e^2 = 0.2$. a. Influence of a pure resistive load, b. Influence of a pure capacitive load.

In order to study the frequency detuning sensibility of the PE-TMD, we analyze the attenuation at the maximal amplitude of $T_1$ as a function of the variation of the primary structure resonant frequency $\omega_1$. The attenuation is calculated by dividing the maximal amplitude of $T_1$ with PE-TMD by the maximal amplitude of the primary structure without TMD as given in equation (11).

$$\text{Gain } \frac{\text{A}}{A_0} = 20\log_{10}\frac{\text{A}}{A_0} = 20\log_{10}\frac{max(|T_1^{\text{with TMD}}|)}{max(|T_1^{\text{without TMD}}|)} \tag{11}$$

Figure 4 represents the Gain $A/A_0$ as a function of the variation of $\omega_1$ for $k_e^2 = 0.3$ and μ = 0.5 %. The response of the optimal mechanical TMD (without piezoelectric effect) is given in Figure 4.a. for comparison, where $\omega_2$ and $Q_{m2}$ are optimally defined to minimize the maximal amplitude of $T_1$ for $\omega_1 = 1$. For the PE-TMD case, $\omega_2$ and $Q_{m2}$ are taken equal to 1 and 90, respectively. The optimal values of $R_l$ for R-tuning, and of $R_l$ and $C_L$ for RC-tuning are determined for each value of $\omega_1^{modified}$. These values are given in  Figure 4.b. and Figure 4.c. respectively.

Figure 4.a. shows that the PE-TMD reach the same attenuation level (gain of -14.6 dB) than the optimal mechanical device at $\omega_1$. This means that the resistive and capacitive load helps to reach a sufficient attenuation level to be at the optimal attenuation of the TMD. When $\omega_1$ is modified, the TMD is not conveniently tuned, and there is an increase in the amplitude of the maximal value of $|T_1|$ for the mechanical TMD and for the PE-TMD without tuning. Resistive and Capacitive tuning (RC-tuning) by adjusting $R_l$ and $C_l$ helps to maintain low values of gain while modifying $\omega_1$. We observe that no capacitive load is necessary to tune the PE-TMD outside a range of frequency ($\omega_1^{modified} < 1.01$ and $\omega_1^{modified} > 1.08$). When a capacitance is used, the tuning helps to obtain the maximal attenuation (gain of -14.6 dB). In this range, the electromechanical coupling of the PE-TMD helps to tune the apparent stiffness and damping through the tuning of the resistive and capacitive loads. Capacitive load is not necessary outside the range, but the tuning is not optimal. An inductive load would be necessary to enable the tuning [6], but their realization is technically limited as discussed in the introduction and in [9]. Moreover, the R-tuning only is not sufficient to minimize the gain in the RC-tuning range (i.e. $\omega_1^{modified} \in [1.01,1.08]$). As an example, for $\omega_1^{modified} = 1.06$, the gain is equal to -14.6 dB with RC-tuning while it is equal to 12.4 dB with R-tuning.

The attenuation level and frequency band of RC-tuning depends on the normalized parameters of the PE-TMD: the electromechanical coupling coefficient $k_e^2$, the mass ratio $\mu$ and the quality factor $Q_{m2}$. The next section discusses the influence of these parameters on the attenuation level and the frequency bandwidth.

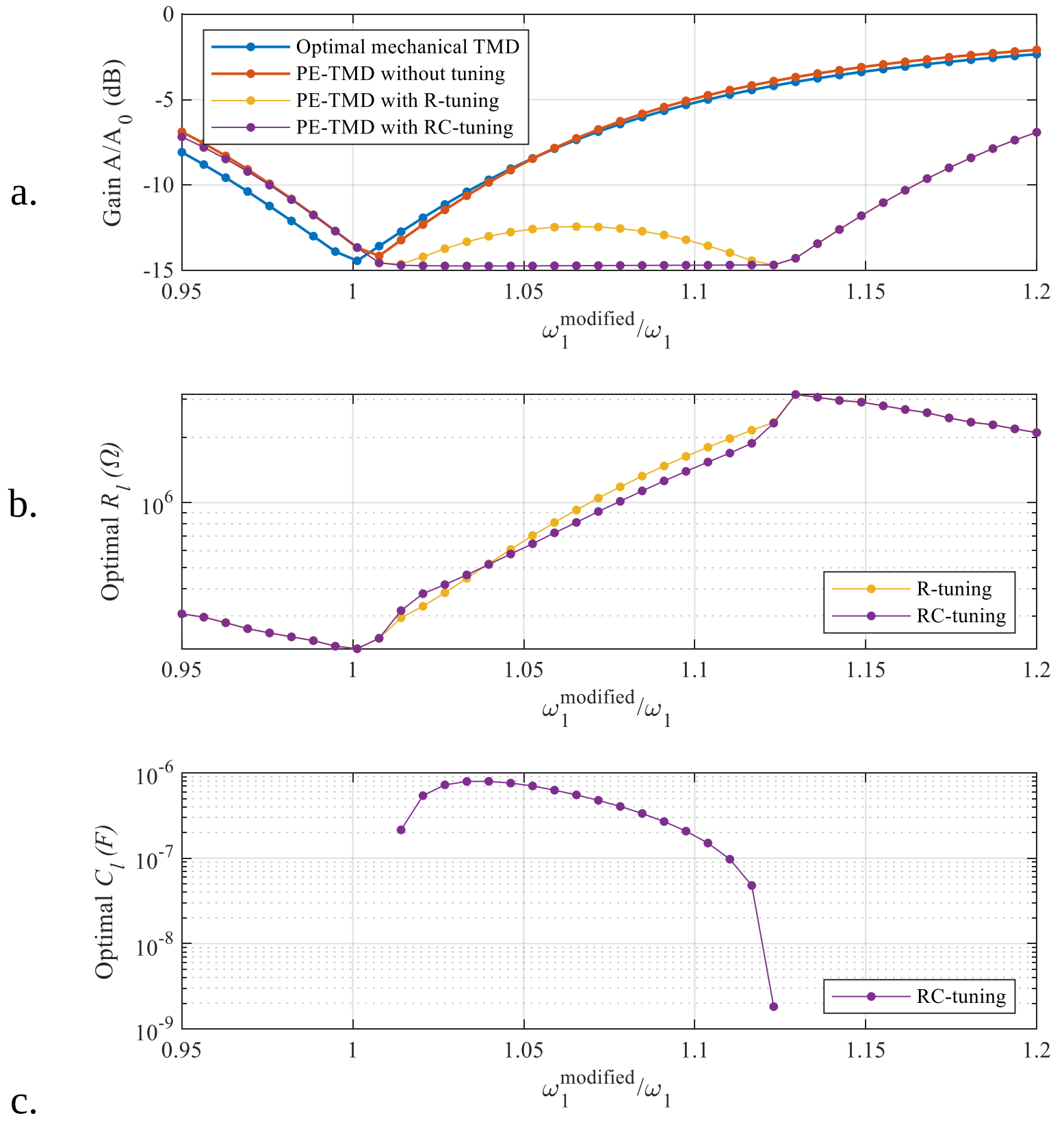

Figure 4 : a. Evolution of the maximal gain $A/A_0$ as a function of $\omega_1$. b. Optimal $R_l$ for R-tuning and RC-tuning with the PE-TMD, c. Optimal $C_l$ for RC-tuning with the PE-TMD. Capacitances are not represented when only R-tuning is sufficient to minimize $A/A_0$

# 4 Parametric studies

## 4.1 Influence on the vibration attenuation

The influence of the three parameters ($k_e^2$, μ and $Q_{m2}$) are studied in this section. They are studied independently with the modification of $\omega_1$ and the adjustment of $R_l$ and $C_l$, while keeping constant the other parameters.

Figure 5 shows the gain of the tuned PE-TMD as a function of the variation of $\omega_1$ for $\mu = 0.5$ % and several values of $k_e^2$. The gain of the optimal mechanical TMD ($K_2$ and $C_2$ selected to minimize the gain at $\omega_1 = 1$) is also represented for comparison. For the low value of $k_e^2$, the induced damping by the electromechanical behavior is not sufficient to reach the same attenuation level as the optimal mechanical TMD. The minimal gain is equal to -11.5 dB and -13.8 dB for $k_e^2 = 0.05$ and $k_e^2 = 0.1$, respectively. We therefore observe that increasing $k_e^2$ helps to maximize the attenuation, to the limit that correspond to the optimal mechanical TMD. Beyond the threshold $k_e^2$, increasing $k_e^2$ increases the frequency band of tunability using resistive and capacitive loads. As a matter of example, the gain is equal to 14.6 dB for $k_e^2 = 0.4$ for $\omega_1^{modified}$ in the range $[1.01; 1.17]$ with RC-tuning, while it is for the range $[1.01; 1.08]$ for $k_e^2 = 0.2$. Moreover notice that only R- tuning is necessary for low and moderate values of $k_e^2$ from Figure 5.b. and c.

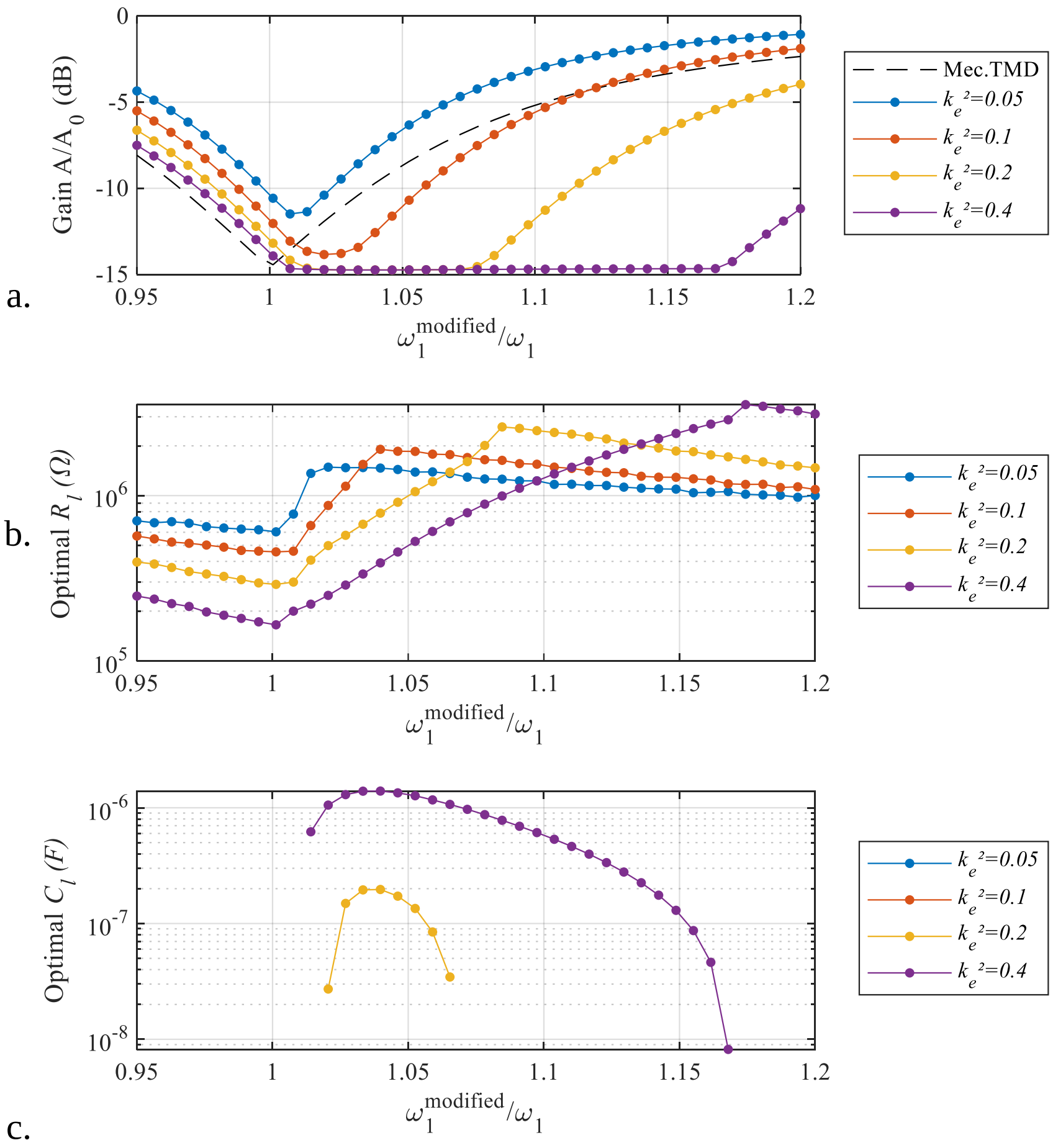

Figure 5 : a. Evolution of the maximal gain $A/A_0$ with the variation of $k_e^2$. b. Optimal resistive load $R_l$, c. Optimal capacitive load $C_l$. Capacitances are not represented when only R-tuning is sufficient to minimize $A/A_0$

Figure 6 shows the gain of the tuned PE-TMD as a function of the variation of $\omega_1$ for $k_e^2 = 0.2$ and several values of μ. The gain of the optimal mechanical TMD ($K_2$ and $C_2$ selected to minimize the gain at $\omega_1 = 1$) is represented for each value of μ. From Figure 6.a., we observe that the frequency range of tunability with RC-tuning is affected by the value of μ. The RC-tuning is feasible and helps to reach the optimal mechanical TMD for $\omega_1^{modified}$ in the range $[1.01, 1.09]$ and $[1.01, 1.08]$ for μ = 0.1 % and μ = 0.5 %, respectively. When μ increases, the electromechanical influence of the PE-TMD decreases. For μ = 1 % and μ = 10 %, the induced damping by the electrical loads does not allow them to reach the mechanical damping of the optimal TMD. Only the R-tuning is necessary to tune the resonant frequency of the TMD. This reveals a challenge in autonomous PE-TMD for the frequency tuning as a stronger $k_e^2$ would be necessary to allow frequency tunability with large value of μ.

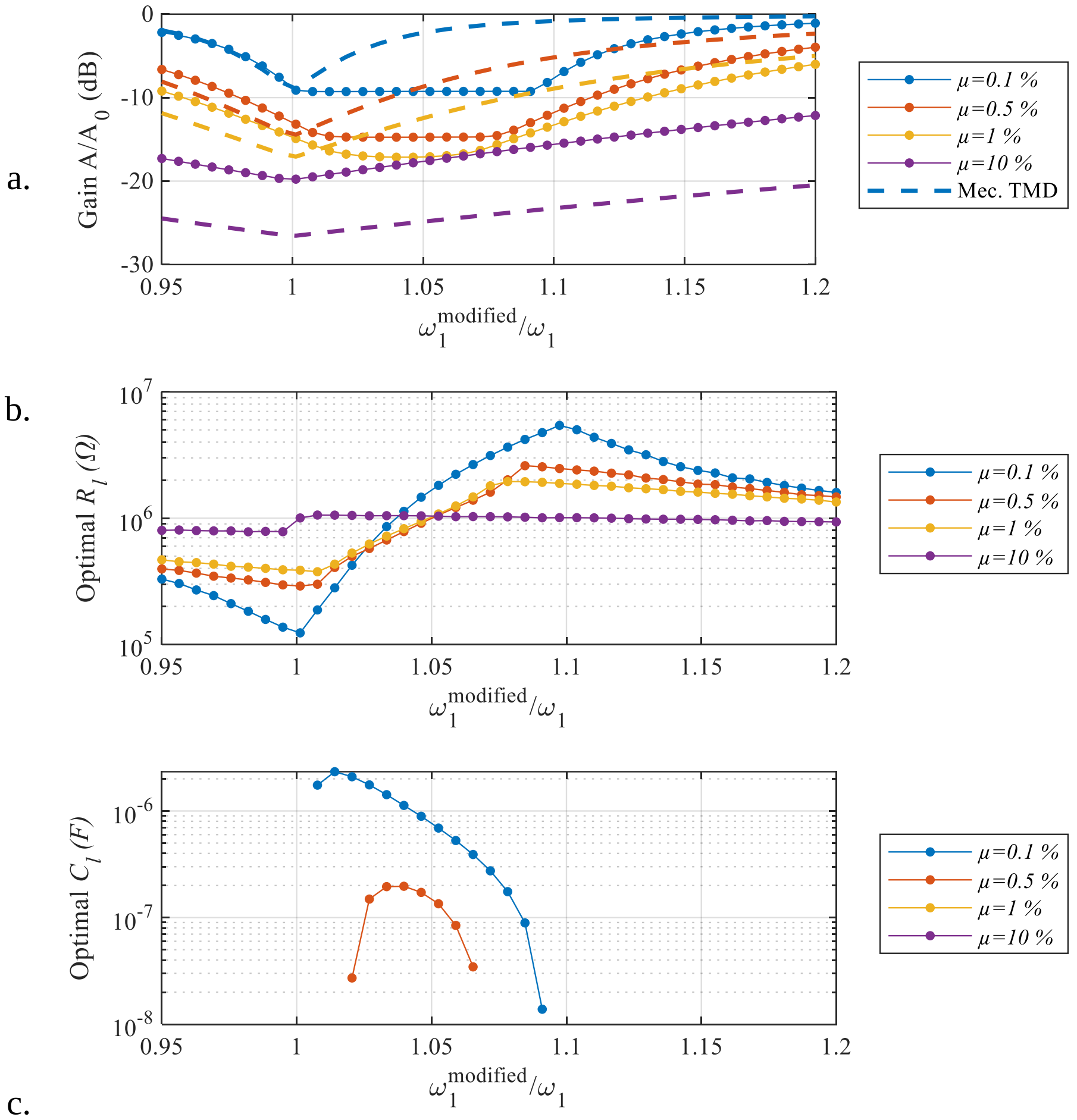

Figure 6 : a. Evolution of the maximal gain $A/A_0$ with the variation of μ. b. Optimal resistive load $R_l$, c. Optimal capacitive load $C_l$. Capacitances are not represented when only R-tuning is sufficient to minimize $A/A_0$

The influence of the mechanical quality factor $Q_{m2}$ on the attenuation performance is illustrated in Figure 7, which shows the gain of the tuned PE-TMD as a function of the variation in the primary structure's natural frequency $\omega_1$ for $k_e^2$= 0.2 and several values of $Q_{m2}$. The results demonstrate that the PE-TMD achieves a consistent attenuation of -14.6 dB for the primary structure's velocity amplitude. The frequency range over which this attenuation is maintained depends on $Q_{m2}$. For $Q_{m2} = 12$, the effective attenuation range spans $\omega_1$ ∈ [1.01, 1.09], while for $Q_m$ = 200, the range narrows to $\omega_1$ ∈ [1.02, 1.07]. This indicates that increasing the mechanical quality factor slightly reduces the bandwidth of effective attenuation, even though the maximum attenuation level remains unchanged. The electromechanical behavior of the PE-TMD helps to reach to the optimal damping level.

a.

b.

c.

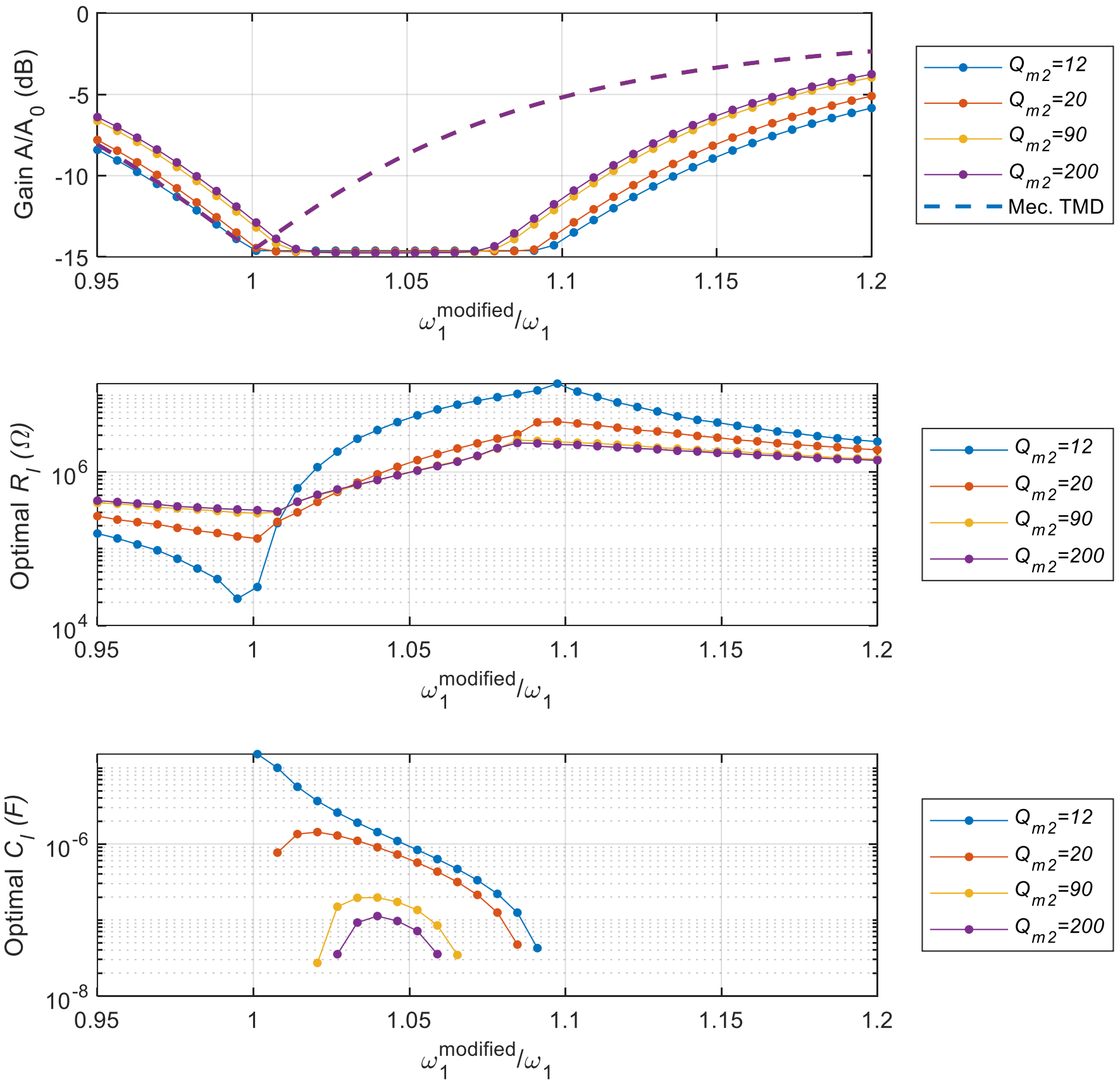

Figure 7 : a. Evolution of the maximal gain $A/A_0$ with the variation of $Q_{m2}$. b. Optimal resistive load $R_l$, c. Optimal capacitive load $C_l$.

## 4.2 Influence on the harvested power and relative velocity of the PE-TMD

In this section, we study the harvested power by the PE-TMD as a function of the parameters μ, $k_e^2$ and $Q_m$. The electrical loads $R_l$ and $C_l$ are tuned to maximize the attenuation, as previously described in Section 0. As the power is related to the relative displacement $x_2 - x_1$ of the PE-TMD, the relative velocity $\dot{x_2} - \dot{x_1}$ is also studied to provide a comprehensive understanding of the system's behavior. The mean power normalized to the square of the input force is computed using the voltage amplitude and the resistive load as expressed in (12). As the capacitance involve an energy storage, only the energy that flows in the resistive loads represents energy dissipation that can be harvested. The relative displacement of the PE-TMD is studied using the transfer function $T_2$ given in (13) and its amplitude is compared to the maximal amplitude of the primary structure without TMD as represented in the gain in (14).

$$\text{Normalized Power} = \frac{P}{F^2} = \frac{1}{F^2}\frac{|\tilde{V}|^2}{2R_l} \tag{12}$$

$$T_2(\omega) = \frac{j\omega(\widetilde{X_2} - \widetilde{X_1})}{\tilde{F}} \tag{13}$$

$$\text{Gain} = 20\log_{10}\frac{A_2}{A_0} = 20\log_{10}\frac{|T_2|_{\omega=\omega_{@\max|T_1|}}}{max(|T_1^{\text{without TMD}}|)} \tag{14}$$

Figure 8 shows the normalized mean power by the PE-TMD for the various values of coupling coefficient ${k_e}^2$, mass ratio μ, and quality factor $Q_{m2}$, as a function of $\omega_1^{modified}$ for the configurations ($R_l, C_l$ and $\omega$) considered in Figure 5 to Figure 7. Figure 9 shows the corresponding relative velocities computed from (14).

In the frequency range where $R_l$ and $C_l$ are tuned to achieve the lowest gain ($A/A_0$), the harvested power is minimized, as shown in Figure 8.a. This behavior is consistent with Figure 9.a, which demonstrates that the relative velocity $A_2/A_0$ is limited by the equivalent damping of the optimal TMD. Consequently, the harvested power remains constrained within this tuned frequency range. Conversely, outside the RC-tuning frequency range, both the harvested power and the relative velocity increase significantly. This indicates that while the PE-TMD achieves optimal attenuation within the tuned range, the relative displacement, and therefore the potential for energy harvesting, becomes more pronounced outside this range.

The relative velocity of the PE-TMD is significant, reaching +12 dB compared to the maximal velocity of the primary structure at resonance. This presents optimization challenges for the TMD, as piezoelectric systems can be brittle and may not withstand such large displacements. Increasing $k_e^2$ has limited impact on the harvested power, in the same way as observations in conventional energy harvesting applications for $k_e^2 Q_m > 2$. However, increasing $k_e^2$ effectively reduces the amplitude of the relative displacement $T_2$ as the gain $A/A_0$ decreases in Figure 5. This shows that while a higher coupling coefficient improves vibration attenuation, it does not necessarily lead to greater energy harvesting.

The mass ratio µ significantly influences the system's behavior in Figure 8.b and Figure 9.a. Increasing µ reduces the harvested power, as a larger PE-TMD mass requires smaller displacements to achieve equivalent kinetic energy and effective dissipation. For small values of µ, the relative displacement of the PE-TMD can be important. As an example, the PE-TMD relative velocity can reach +17 dB compared to the primary structure alone for µ=0.1%.

While the influence of the quality factor $Q_{m2}$ is not significant on the vibration attenuation, it strongly impacts the harvested power as shown by Figure 8.c. Increasing $Q_{m2}$ leads to higher power output, as less energy is dissipated in mechanical losses, and can be harvested. Meanwhile, both the gains $A/A_0$ and the relative displacement gain $A_2/A_0$ are almost independent of $Q_\mathrm{m}$. This indicates that a higher quality factor improves electrical power generation without affecting the TMD's attenuation performance.

Table 2 summarizes the contributions of each parameter to the desired performance. It is clear that the piezoelectric coupling coefficient must be maximized in all aspects. However, the table also illustrates the complexity of achieving a good level of attenuation and a wide range of frequency tunability. Indeed, a good mass ratio is necessary for damping and reduces the relative displacement of the PE-TMD, but it hinders energy harvesting and frequency tuning.

Table 2: Influence of the parameters on the performance. ↗ : Should be maximized, ↘ : Should be minimized, $=$ : No influence, ↱ : Should me maximized till a specific value

| | Coupling coefficient $k_e^2$ | Mass ratio µ | Quality factor $Q_m$ |
|---|---|---|---|
| Attenuation level | ↱ | ↗ | $=$ |
| Frequency tunability | ↗ | ↘ | ↘ |
| Harvested power | ↱ | ↘ | ↗ |
| Relative displacement of the PE-TMD | ↱ | ↗ | $=$ |

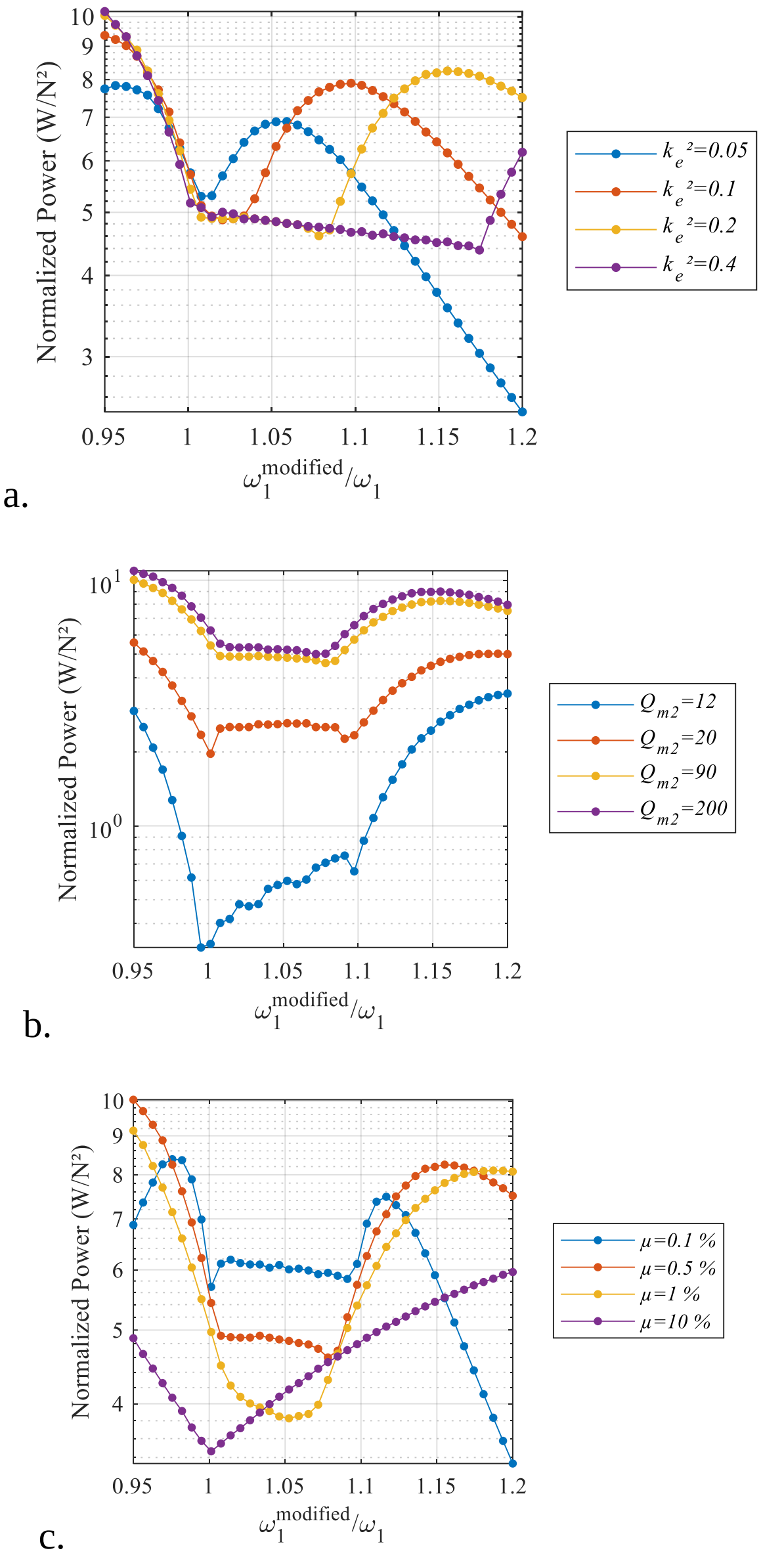

Figure 8 : Normalized mean power for several values of a. coupling coefficient $k_e{}^2$, b. mass ratio μ, and c. quality factor $Q_m$. When they are not specified, the value of $k_e^2$, μ and $Q_m$ are 0.2, 0.5 % and 90 respectively.

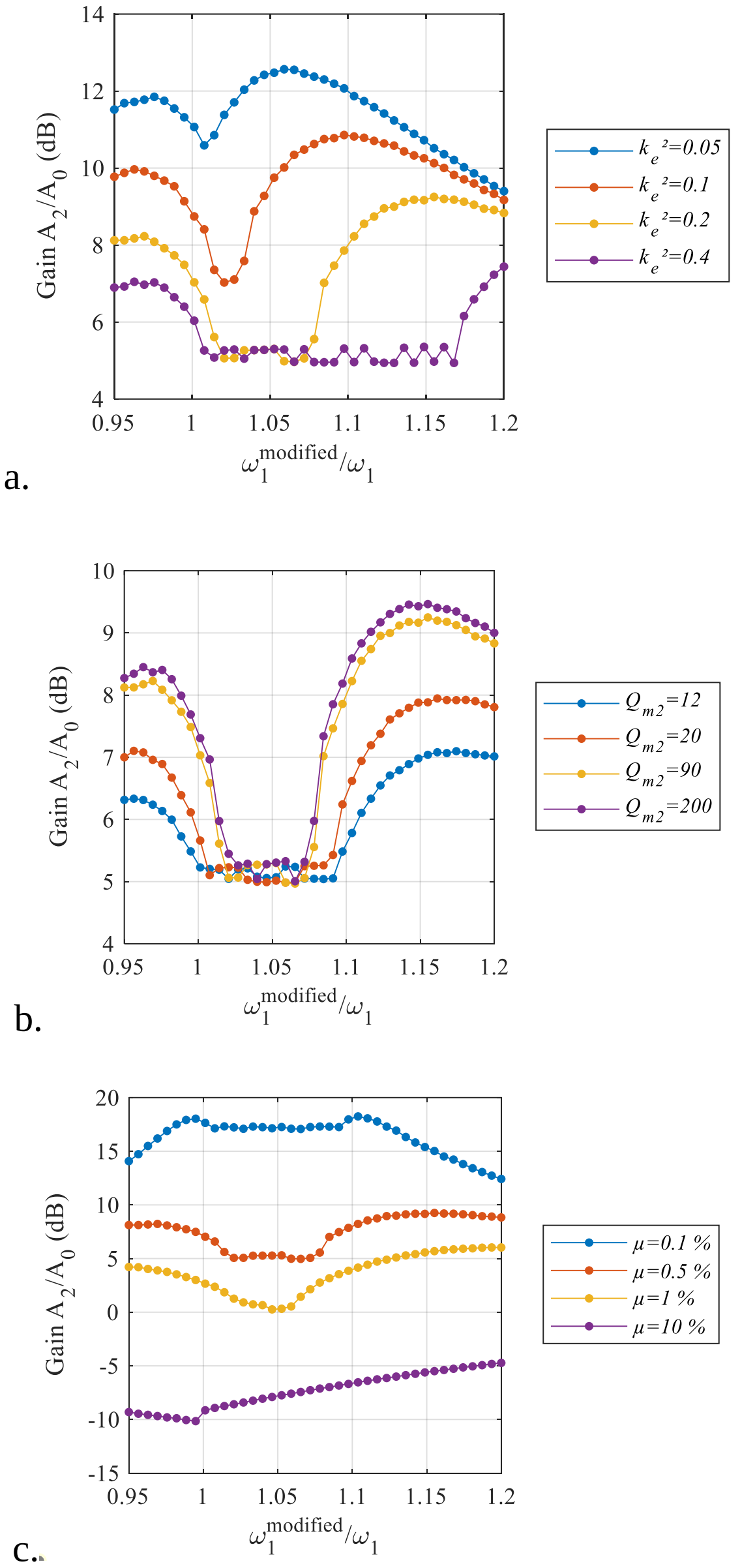

Figure 9 : Relative velocity of the PE-TMD for several values of a. coupling coefficient $k_e{}^2$, b. mass ratio μ, and quality factor $Q_m$. When they are not specified, the value of $k_e^2$, μ and $Q_m$ are 0.2, 0.5 % and 90 respectively.

# 5 Conclusion

This work presents a comprehensive analytical study of self-powered piezoelectric tuned mass dampers (PE-TMDs), demonstrating their potential for autonomous frequency tuning and vibration attenuation through resistive and capacitive load adjustment (R-tuning and RC-tuning). The developed model confirms that PE-TMDs can achieve optimal attenuation levels, comparable to mechanical TMDs, while offering frequency tunability over significant ranges.

Key findings are the following:

- In order to enable frequency tuning, PE-TMD with strong $k_e^2$ is necessary, but its influence is limited when μ increases. Increasing μ necessitates higher values of $k_e^2$ to keep interesting the frequency tuning capability.
- For moderate $ke^2$ and μ, it is not necessary to use capacitive loads; the R tuning is sufficient
- The high amplitude displacement of the PE-TMD can be a technical limitation, especially for low μ values.
- Decreasing the quality factor of PE-TMD increases the frequency range of tuning but decreases the harvested power

Future work will focus on experimental validation, to be presented at the conference, and the study of nonlinear electrical techniques to further enhance performance.

## Acknowledgements

The authors would like to acknowledge of Amaury Gegout its preliminary study during its internship and to Blaise Girard for the future manufacturing of the prototypes. This work was supported by the French National Agency for research (ANR) through the project ANR-25-CE51-6347 (ANR FAPARVA).